\documentstyle{article}

\begin{document}

\begin{center}

{\large {\bf Correlation between the quenching of total GT$_+$ strength
and the increase of E2 strength}}

\vspace{0.2in}

N. Auerbach$^{(1)}$, D. C. Zheng$^{(2)}$, L. Zamick$^{(3)}$,
	and B. A. Brown$^{(4)}$

\end{center}

\thispagestyle{empty}

\vspace{0.2in}
\begin{small}

\hspace{-0.4in}{\it $^{(1)}$School of Physics and Astronomy,
		Tel--Aviv University, Tel Aviv, 69978, Israel}

\hspace{-0.4in}{\it $^{(2)}$Department of Physics, University of Arizona,
	Tucson, AZ 85721}

\hspace{-0.4in}{\it $^{(3)}$Department of Physics and Astronomy,
	Rutgers University, Piscataway, NJ 08855}

\hspace{-0.4in}{\it $^{(4)}$Department of Physics and Astronomy,
	Michigan State University, East Lansing, MI 48824}

\end{small}

\vspace{0.5in}

\begin{abstract}
Relations between the total $\beta_+$
Gamow-Teller (GT$_+$) strength and the E2 strength
are further examined. It is found that in shell-model calculations for
$N$=$Z$ nuclei, in which changes in deformation are induced by varying
the single-particle energies, the total GT$_+$ or GT$_-$ strength
decreases monotonically with increasing values of the B(E2)
from the ground state to the first excited $J^{\pi}$=$2^{+}$ state.
Similar trends are also seen for the double GT transition amplitude
(with some exceptions) and for the spin part of the total M1 strength as a
function of B(E2).
\end{abstract}

\pagebreak

\section{INTRODUCTION}
In Ref.\cite{auerbach} a relation between the B(E2) value from the ground
state to the
first excited $J^{\pi}$=$2^{+}$ state and the quenching of Gamow-Teller
(GT) strength has been pointed out. The quenching we will discuss here is
caused by those nuclear structure effects that do {\it not} affect
the sum rule:
\begin{equation}
B_t({\rm GT}_-) - B_t({\rm GT}_+)  = 3(N-Z),
\end{equation}
where $B_t({\rm GT}_-)$ and $B_t({\rm GT}_+)$ are the $(p,n)$ $\beta_-$ and
$(n,p)$ $\beta_+$ GT transition strengths. (We will refer to the quenching
of the sum rule value as the quenching due to the renormalization of the
GT operator.)

The reduction of the total GT strength, discussed here,
is such that it affects the total GT$_+$ and GT$_-$ strengths
in the {\it same additive} way and, therefore, cancels out when the
difference $B_t({\rm GT}_-) - B_t({\rm GT}_+)$ is taken.
Because of the Pauli blocking, in nuclei with a neutron excess,
the $\beta_{+}$
branch ($\sigma t_+$) of a GT transition is usually much weaker than
the $\beta_-$ branch. Hence, when the above mentioned quenching occurs,
it is more pronounced in the $\beta_+$ transition.

The quenching of the total GT$_+$
strength, which concerns us, is due to the presence of multi-particle,
multi-hole components in the wave function of the initial and final states,
i.e. $np$-$nh$ with $n>2$. In a deformed nucleus, the amount of
$np$--$nh$ admixture is large, so that the $2p$--$2h$ space is
not sufficient to describe the ground state or transitions from the ground
state to the GT states.
A full shell-model calculation in an extended space is expected to
predict the amount of quenching of the GT strength \cite{bab}.
However, the possibility of performing
such extended shell-model calculations is limited to
very few cases. Various approximation schemes, such as the RPA and, more
widely used in this context, the QRPA, describe well only the
$2p$-$2h$ part of the $np$-$nh$ space. Consequently,
It is important to try and develop a procedure that would enable us
to predict the quenching
of the GT strength, by finding a relationship of the
GT strength to some other observable
that is more easily accessible experimentally or theoretically.

Deformed nuclei have large B(E2) values. This consideration should provide
a simple indication of possible correlations between the quenching of the GT
strength and the B(E2) values.
In Ref.\cite{auerbach}, such a relationship
was suggested and tested for the case
of $^{26}\mbox{Mg}$, for which a full $sd$-space shell-model calculation
is still possible. It was found that the total GT$_+$
strength is a monotonically
descending function of the B(E2) value for the transition
from the ground state to the first excited
$J^{\pi}=2^{+}$ state of the parent nucleus.
This relation was then used in Ref.\cite{auerbach} to estimate the total
GT strength in the $^{54}\mbox{Fe}$ region.

In the present paper, we extend these calculations
to more cases where one is able to perform large-space shell-model
calculations of the total GT strength. The purpose is to confirm
the previous conjecture about the above relationship between
the total GT$_+$ or
GT$_-$ strength and the B(E2) value,
to provide more insight into the origin and to set
some limits on the application of this relationship.

\section{THE CALCULATION}

Two of the few nuclei that are available for a complete space
calculation are $^{20}\mbox{Ne}$ and $^{44}\mbox{Ti}$. For these
two nuclei, the GT$_-$ and GT$_+$ strengths are equal and, therefore,
the quenching will affect the two branches to the same degree.
In $^{20}\mbox{Ne}$,
we allow the two valence protons and two valence neutrons
to occupy the complete $sd$ shell ($d_{5/2}$, $s_{1/2}$ and $d_{3/2}$),
while in $^{44}\mbox{Ti}$ the two valence protons and two valence neutrons
are allowed to occupy the complete $fp$ shell
($f_{7/2}$, $p_{3/2}$, $f_{5/2}$ and $p_{1/2}$).
The calculations are performed with two-body interactions used in
previous calculations
to describe the structure of these nuclei. In
$^{20}\mbox{Ne}$ we use the Wildenthal interaction \cite{wildenthal},
while in $^{44}\mbox{Ti}$ two sets of matrix elements are used,
the modified renormalized Kuo-Brown (MKB) interaction \cite{mkb} and
the FPD6 interaction \cite{fpd6}.
As for single-particle energies, we use, in
$^{20}\mbox{Ne}$, the set (given in MeV):
\begin{equation}
\epsilon_{d_{5/2}} = 0.0, \hspace{0.2in}
  \epsilon_{s_{1/2}} = 0.78(1+x), \hspace{0.2in}
  \epsilon_{d_{3/2}} = 5.59(1+x)
\end{equation}
and in $^{44}\mbox{Ti}$ we use, with the MKB:
\begin{equation}
\epsilon_{f_{5/2}} = 0.0, \hspace{0.2in}
  \epsilon_{p_{3/2}} = 2.1(1+x), \hspace{0.2in}
  \epsilon_{f_{5/2}} = 4.4(1+x), \hspace{0.2in}
  \epsilon_{p_{1/2}} = 8.2(1+x);
\end{equation}
and, with the FPD6:
\begin{equation}
\epsilon_{f_{5/2}} = 0.0, \hspace{0.2in}
  \epsilon_{p_{3/2}} = 1.89(1+x), \hspace{0.2in}
  \epsilon_{f_{5/2}} = 3.91(1+x), \hspace{0.2in}
  \epsilon_{p_{1/2}} = 6.49(1+x).
\end{equation}
The variable $x$ is a parameter that will enable us to vary the single-particle
spacing and, in particular, the spin-orbit splitting. In this manner
we are also able to change the nuclear deformation.
The value $x$=0 corresponds to the realistic case used with the above
two-body interactions to reproduce the empirical properties of the
nuclei under discussion.

It is also instructive to look at the SU(3) limit \cite{elliot}
for these nuclei, by constructing,
in this limit, their wave functions and computing their
B(E2) values. In the SU(3) limit, the total GT$_+$ and GT$_-$ strengths
from the ground state of an $N$=$Z$ nucleus vanish.

In the SU(3) model, the spin-orbit splitting is put to zero,
so that the spin-orbit
partners are degenerate. The spacing between the single-particle
states that are not spin-orbit partners are non-zero
but have definite values, as given by the diagonal matrix elements
of the Elliott, say, quadrupole-quadrupole
(Q$\cdot$Q) interaction.

\section{RESULTS}

Our calculated results are given in Table I for
$^{20}\mbox{Ne}$ and in Table II for $^{44}\mbox{Ti}$.
In addition to the total GT$_+$ strength B$_t$(GT$_+$),
which we will discuss first,
the results for many other quantities for different values
of the spin-orbit splitting parameter $x$ are also given in the tables.

In the single $j$-shell model, in both
$^{20}\mbox{Ne}$ and $^{44}\mbox{Ti}$, the total GT$_+$ strength
B$_t$(GT$_+$)
is equal to 6.0. As configuration mixing
is introduced, B$_t$(GT$_+$) is reduced.
In the realistic case (i.e. $x$=0), it is reduced
to about 0.55 in $^{20}\mbox{Ne}$ and to 1.88 with the MKB
(1.27 with the FPD6) in $^{44}\mbox{Ti}$.

As the parameter $x$ is increased,
the spacing between the lowest single-particle state
($d_{5/2}$ in $^{20}\mbox{Ne}$ and $f_{7/2}$ in $^{44}\mbox{Ti}$)
and other single-particle states is increased and the states approach the
limit of the pure configurations. The B(E2) values to the first excited
$J^{\pi}=2^{+}$ state B$_1$(E2) decrease, and, at the same time, the total
GT$_+$ strength B$_t$(GT$_+$) increases monotonically.
This monotonic behavior of
the B$_t$(GT$_+$) values as a function of B$_1$(E2) is also shown in Fig.1.
The only slight deviation
from a monotonic behavior occurs in
$^{20}\mbox{Ne}$ for the value of $x$=--$1$ which corresponds
to complete degeneracy in the single-particle spectrum.
For comparison, in both tables and in Fig.1,
the B$_1$(E2) values for in the SU(3) model are also shown.
As remarked earlier, in this limit, the total GT$_+$ strength is zero.

Note that the GT$_+$ strength from the ground state
in $^{44}\mbox{Ti}$ to the lowest $J^{\pi}$=$1^+$ state in
$^{44}\mbox{V}$, B$_1$(GT$_+$), does not show the same monotonic behavior
as the total strength does. We can see from Table II that the change
in the behavior of B$_1$(GT$_+$) occurs around the point where the quadrupole
moment $Q$ changes sign and the nuclear shape changes from prolate
to oblate.

Now we discuss the results for the ground state to ground state
($J^{\pi}=0^{+},\; T=2 \rightarrow J^{\pi}=0^{+},\; T=0$)
double Gamow-Teller (DGT) transition amplitude, A$_1$(DGT).
This is of interest to double-beta-decay
calculations and possibly to double-charge-exchange reactions with pions.
The A$_1$(DGT) values for different choices of $x$ are given in the
rightmost column in Tables I and II.
In Fig.2, we show the results for A$_1$(DGT) as a function of
B$_1$(E2). We see that when the Wildenthal  \cite{wildenthal}
and the FPD6 \cite{fpd6} interactions
are used for $^{20}\mbox{Ne}$ and $^{44}\mbox{Ti}$, respectively,
the DGT amplitude A$_1$(DGT) decreases monotonically with
increasing B$_1$(E2).
This has been noted previously in Ref.\cite{zheng}. In this
reference, we also found that when the MKB interaction is used,
A$_1$(DGT) as a function of the single particle splittings
deviates slightly from the monotonic behavior for large values
of $x$ or small values of B$_1$(E2). One can also see this
non-monotonic behavior from Table II and Fig.2.

Note that the DGT amplitude A$_1$(DGT) vanishes in both the
SU(3) limit and the SU(4) limit,
due to the vanishing of the total GT$_+$ and
GT$_-$ strengths from the final $N$=$Z$ nucleus.

Finally we discuss the results for the total M1 strength.
In Tables I and II, we give the total spin, orbital and full
M1 strengths [B$_s$(M1), B$_o$(M1) and B$_t$(M1)]
for different values of $x$. The M1 strength to the lowest
$J^{\pi}$=$1^+$, $T$=1 state, B$_1$(M1), is also shown.
{}From Fig.3, in which we plot the total spin and orbital M1 strengths
as a function of B$_1$(E2), it is evident that the total spin
M1 strength decreases monotonically with increasing B$_1$(E2), while
the total orbital
M1 strength increases monotonically with increasing B$_1$(E2).
The strong correlation between the total {\it orbital} M1 strength
and the B$_1$(E2) values has recently been well established
both experimentally \cite{bohle,richter} and theoretically
\cite{loi,iach,heyde,zamick}.
That the total spin M1 strength decreases with increasing
nuclear deformation has also been noted previously in
Refs.\cite{zamick,zzm}. It was pointed out in Ref.\cite{zzm} that
in the large deformation limit, the spin M1 strength vanishes.
Note that the B$_1$(M1) values in $^{44}\mbox{Ti}$, which are for
the transition to the lowest $J^{\pi}$=$1^+$ state alone, are not monotonic
with increasing B$_1$(E2) (see Table II).

\section{FURTHER DISCUSSIONS}

Let us return to the total GT$_+$ strength. The decrease of the
total GT$_+$ strength B$_t$(GT$_+$)
as a function of the increasing B$_1$(E2) values, is quite independent
on the kind of interaction used as can be seen from Table II, where
the results for $^{44}\mbox{Ti}$ are given for two effective interactions,
the MKB \cite{mkb} and the FPD6 \cite{fpd6}.
We can also see this from Fig.1, in which
the two curves for $^{44}\mbox{Ti}$ and the curve for
$^{20}\mbox{Ne}$ have very similar shapes.
These curves are also very similar to the curve obtained for the
total GT$_+$ strength in $^{26}\mbox{Mg}$,
as described in Ref.\cite{auerbach}.

We should remark that the dependence of the total GT$_+$ strength on the
B$_1$(E2) values is not a single-valued function. This applies,
in particular, to the small values of the GT$_+$ strength, i.e.,
when the quenching is very large. There  are many models in which
the B$_1$(E2) values are different and which will give zero or nearly zero
GT$_+$ values. For example, in the Cartesian Harmonic Oscillator
model for $^{20}\mbox{Ne}$, the ground state wave function is given by
the state in which the excess four nucleons occupy the orbit with
($n_x$=$0$, $n_y$=$0$, $n_z$=$2$). To form a final $J^{\pi}$=$1^{+}$
state, one has to excite one of the nucleons in the intrinsic
state to, say, ($n_x$=$0$, $n_y$=$1$, $n_z$=$1$). The spin
operator cannot change the spatial wave function. Thus the GT or
spin M1 transition matrix elements are zero.
However, the B$_1$(E2) value obtained in this
model is smaller than, for example, the B$_1$(E2) value obtained from the
SU(3) wave functions. Nevertheless, we believe that for larger
values of the GT$_+$ strength, the correspondence between a given B$_1$(E2)
value and the total GT$_+$ strength is better defined and
should provide us with a practical
way to estimate the quenching of the GT$_+$ strength from the measured
B$_1$(E2) values.
We base this conclusion on the fact that the curves shown in Fig.1 for
$^{20}\mbox{Ne}$ and $^{44}\mbox{Ti}$, as well as the results given in
Ref.\cite{auerbach}, show a very similar, quite ``universal'' behavior.

We emphasize again that the quenching of the total
GT$_+$ strength that we are addressing here applies also to the total
GT$_-$ strength in the same additive way. However, since most nuclei have
a neutron excess and the total GT$_-$ strength is usually much larger than the
total GT$_+$ strength, our results are more useful in estimating the
quenching of the GT$_+$ strength in $\beta_{+}$ transitions.

Recently the total GT$_+$ strength was measured in the $(n,p)$ reaction
on $^{54}\mbox{Fe}$ and $^{56}\mbox{Fe}$ \cite{ronn}.
The total GT$_+$ strength found for $^{54}\mbox{Fe}$
is $B_t({\rm GT}_+)$=3.5, while for $^{56}\mbox{Fe}$,
$B_t({\rm GT}_+)$=2.3.
The authors point out that these results
are in agreement with the findings of reference \cite{auerbach}. The
B$_1$(E2) value for $^{56}\mbox{Fe}$ is 620 $e^2 {\rm fm}^4$,
larger than the B$_1$(E2) value of 980 $e^2 {\rm fm}^4$
for $^{54}\mbox{Fe}$ \cite{raman}
and, therefore, according to Ref.\cite{auerbach}
and the present work, the quenching of the total GT$_+$ strength should
be larger in $^{56}\mbox{Fe}$ than in $^{54}\mbox{Fe}$.
Indeed, the product of
the B$_1$(E2) value and the total GT$_+$ strength from experiments
is about the same
for these two nuclei (2170 $e^2 {\rm fm}^4$ for $^{54}\mbox{Fe}$
and 2254 $e^2 {\rm fm}^4$ for $^{56}\mbox{Fe}$).
This recent experimental work provides a nice example of the
kind of use one can make
of the relationship
we have established in this work and in Ref.\cite{auerbach}.

\vspace{0.3in}

We thank G. F. Bertsch for useful discussions and
B. R. Barrett for a careful reading of the manuscript
and discussions. One of us (D.C.Z.)
thanks the Institute for Nuclear Theory at the University of Washington
for its hospitality and support during his visit.
This work was supported in part by the National Science Foundation under
Grant No. PHY-90-17077 and Grant No. PHY-9103011
and by the Department of Energy under contract number
DE-FG05-86ER-40299.

\vspace{0.2in}

\begin{small}

\pagebreak

{\bf Table I}.
The GT$_+$, M1 and E2 transition strengths
from the ground state in ${}^{20}\mbox{Ne}$ as a function of $x$,
a parameter signifying the splitting between the single-particle
energy of the $d_{5/2}$ orbit and those of the $s_{1/2}$ and
$d_{3/2}$ orbits (see text for more details).
The calculations were performed in the full $sd$ space, using
the two--body matrix elements of the Wildenthal interaction \cite{wildenthal}.
In the table, $Q$ is the quadrupole moment of
the $J^{\pi}$=$2^{+}_1$, $T$=0 state
and $A_1(DGT)$ is the ground state to ground state DGT amplitude.
M1 strengths are in units of $\mu_N^2$ and
E2 strengths are in units of $e^2 {\rm fm}^4$.

\begin{center}

\begin{tabular}{r|ccccccccc} \hline\hline
$x$\hspace{0.05in} & $B_{t}(GT_+)^{a}$ & $B_{s}(M1)^{b}$ & $B_{o}(M1)^{c}$
    & $B_{t}(M1)^{d}$ & $B_1(GT_+)^{e}$ & $B_1(M1)^{f}$
    & $B_1(E2)^{g}$ & $Q^{h}$ & $A_{1}(DGT)^{i}$ \\ \hline
 -1.0&0.116&0.305&1.487&1.781&0.000&0.566&298.4&-15.78&-0.051 \\
 -0.5&0.235&0.620&1.356&1.942&0.050&1.363&307.2&-16.07&-0.025 \\
  0.0&0.548&1.449&1.203&2.566&0.143&1.962&303.0&-15.81& 0.246 \\
  0.5&0.949&2.508&1.075&3.439&0.241&2.412&291.2&-15.21& 0.486 \\
  1.0&1.360&3.595&0.973&4.375&0.332&2.755&276.2&-14.36& 0.702 \\
  1.5&1.739&4.596&0.889&5.255&0.409&3.007&260.7&-13.27& 0.883 \\
  2.0&2.065&5.460&0.819&6.025&0.471&3.182&245.6&-11.97& 1.028 \\
  4.0&2.880&7.612&0.643&7.967&0.589&3.427&197.7&-5.920& 1.350 \\
  8.0&3.423&9.050&0.520&9.290&0.631&3.450&153.1& 1.108& 1.525 \\ \hline
SU(3)&0.000&0.000&1.091&1.091&0.000&1.091&346.3&-17.07& 0.000 \\ \hline\hline
\end{tabular}

\end{center}

$^{a}B_t(GT_+)$: Total GT$_+$ strength to $J^{\pi}$=$1^+$, $T$=1 states;

$^{b}B_s(M1)$: Total spin M1 strength to $J^{\pi}$=$1^+$, $T$=1 states;

$^{c}B_o(M1)$: Total orbital M1 strength to $J^{\pi}$=$1^+$, $T$=1 states;

$^{d}B_t(M1)$: Total M1 strength to $J^{\pi}$=$1^+$, $T$=1 states;

$^{e}B_1(GT_+)$: GT$_+$ strength to the lowest $J^{\pi}$=$1^{+}$, $T$=1 state;

$^{f}B_1(M1)$: M1 strength to the lowest $J^{\pi}$=$1^{+}$, $T$=1 state;

$^{g}B_1(E2)$: E2 strength to the lowest $J^{\pi}$=$2^{+}$, $T$=0 state;

$^{h}Q$: Quadrupole moment of the lowest $J^{\pi}$=$2^{+}$, $T$=0 state;

$^{i}A_{1}(DGT)$: DGT amplitude for the transition
$^{20}\mbox{O} (J^{\pi}=0^{+}_{1},\; T=2) \rightarrow
 ^{20}\mbox{Ne}(J^{\pi}=0^{+}_{1},\; T=0)$.

\pagebreak

{\bf Table II}.
Same as Table I but for the transitions from
the ground state in ${}^{44}\mbox{Ti}$.
The calculations were performed in the full $fp$ space using
the two--body matrix elements of the MKB \cite{mkb}
and of the FPD6 \cite{fpd6} interactions.

\begin{center}

\begin{tabular}{c|r|ccccccccc} \hline\hline
Int. &$x$&$B_{t}(GT_+)$&$B_{s}(M1)$&$B_{o}(M1)$&$B_{t}(M1)$
	& $B_{1}(GT_+)$&$B_{1}(M1)$&$B_1(E2)$&$Q$&$A_1(DGT)$ \\ \hline
MKB&-1.0&0.095&0.252&2.713&2.985&0.006&0.551&829.3&-25.99& 0.220 \\
   &-0.5&0.723&1.911&2.266&4.134&0.289&3.097&782.7&-24.61& 0.679 \\
   & 0.0&1.884&4.981&1.661&6.487&0.509&3.902&624.8&-14.97& 1.103 \\
   & 0.5&2.667&7.051&1.240&8.087&0.529&3.706&484.4&-0.923& 1.257 \\
   & 1.0&3.068&8.111&1.030&8.928&0.521&3.583&414.2& 6.043& 1.287 \\
   & 1.5&3.289&8.696&0.921&9.409&0.514&3.522&382.4& 9.038& 1.282 \\
   & 2.0&3.416&9.058&0.859&9.717&0.509&3.488&365.9& 10.56& 1.269 \\
   & 4.0&3.672&9.706&0.762&10.30&0.496&3.437&341.9& 12.71& 1.222 \\
								  \hline
FPD6&-1.0&0.040&0.107&2.881&2.988&0.000&0.401&835.6&-26.40&0.058 \\
   &-0.5&0.376&0.995&2.631&3.596&0.154&2.603&804.4&-25.64& 0.283 \\
   & 0.0&1.269&3.354&2.126&5.416&0.403&3.841&697.8&-21.61& 0.809 \\
   & 0.5&2.125&5.617&1.657&7.219&0.495&3.816&580.5&-13.12& 1.176 \\
   & 1.0&2.677&7.076&1.359&8.407&0.484&3.462&490.6&-4.184& 1.367 \\
   & 1.5&3.010&7.957&1.185&9.143&0.461&3.204&434.3& 1.629& 1.462 \\
   & 2.0&3.222&8.516&1.081&9.621&0.442&3.039&401.1& 4.984& 1.512 \\
   & 4.0&3.603&9.523&0.911&10.52&0.402&2.758&351.1& 9.846& 1.571 \\ \hline
\multicolumn{2}{c|}{SU(3)}
        &0.000&0.000&1.563&1.563&0.000&1.563&1067&-29.78& 0.000 \\
					 \hline\hline
\end{tabular}

\end{center}

\end{small}

\pagebreak

\section*{FIGURE CAPTIONS}

{\bf Fig.1}. The total GT$_+$ or GT$_-$ strength from the ground states
in $^{20}\mbox{Ne}$ (using the Wildenthal interaction)
and in $^{44}\mbox{Ti}$ (using the MKB and the FPD6 interactions)
as a function of the B(E2) values from the ground state to the
first excited $J^{\pi}$=$2^+$ state in the Weisskopf units.
The symbols ``+'', ``$\times$'', ``$\triangle$'' {\it etc.}
signify the points for which actual calculations were performed.
The results in the SU(3) limit are also marked.

\vspace{0.4in}

{\bf Fig.2}. The DGT transition amplitude from the
$J^{\pi}$=$0^+_1$, $T$=2 state to the $J^{\pi}$=$0^+_1$, $T$=0 state
as a function of the B(E2) from the $J^{\pi}$=$0^+_1$, $T$=0 state
to the $J^{\pi}$=$2^+_1$, $T$=0 state.
The Wildenthal interaction was used for
$^{20}\mbox{Ne}$ and the MKB and the FPD6 interactions were used for
$^{44}\mbox{Ti}$.
The symbols ``+'', ``$\times$'', ``$\triangle$'' {\it etc.}
signify the points for which actual calculations were performed.

\vspace{0.4in}

{\bf Fig.3}. The spin and orbital total M1 transition strengths
from the ground states
in $^{20}\mbox{Ne}$ and in $^{44}\mbox{Ti}$ as a function of
the B(E2) values from the ground state to the first excited
$J^{\pi}$=$2^+$ state.
The Wildenthal interaction was used for
$^{20}\mbox{Ne}$ and the FPD6 interaction was used for
$^{44}\mbox{Ti}$.
The symbols ``+'', ``$\times$'', ``$\triangle$'' {\it etc.}
signify the points for which
actual calculations were performed.

\end{document}